\documentclass[aps,prd,amsmath,amssymb,preprintnumbers,twocolumn]{revtex4}
\usepackage[a4paper,left=20mm,right=20mm,top=25mm,bottom=25mm]{geometry}
\usepackage{graphicx}
\usepackage{color}
\usepackage{subfigure}
\usepackage{pxfonts}
\usepackage{fancyhdr}
\usepackage{multirow}
\usepackage{tikz}

\definecolor{rossoCP3}{cmyk}{0,0.88,0.77,0.40}

\begin{document}

%%%%%%%%%%%%%%%%%%%%%%%%%%%%%%%%%%%%%%%%%%%%%%%%%%%%%%%%%%%%%%%%%%%%%%%%%%%%%%%%%%%%%%%%
\title{\Large \color{rossoCP3}  Starobinsky Model in Rainbow Gravity}

\author{Auttakit Chatrabhuti$^{1}$ and Vicharit Yingcharoenrat$^{2}$}
\affiliation{\vspace{3mm} Department of Physics, Faculty of Science, Chulalongkorn University, Bangkok 10330, Thailand\\
$^{1,2}$Email: {\rm auttakit@sc.chula.ac.th, vicharit.yingcharoenrat@gmail.com}}

\author{Phongpichit Channuie$^{3}$}
\affiliation{\vspace{3mm} School of Science, Walailak University, Thasala, Nakhon Si Thammarat, 80160, Thailand\\
$^{3}$Email: {\rm channuie@gmail.com}}
%%%%%%%%%%%%%%%%%%%%%%%%%%%%%%%%%%%%%%%%%%%%%%%%%%%%%%%%%%%%%%%%%%%%%%%%%%%%%%%%%%%%%%%%

\begin{abstract}
In this paper, we study the Starobinsky model of inflation in the context of gravity's rainbow theory. We propose that gravity rainbow functions can be written in the power-law form of the Hubble parameter.  We present a detailed derivation of the spectral index of curvature perturbation and the tensor-to-scalar ratio and compare the predictions of our models with Planck 2015 data. We discover that in order to be consistent with Planck data up to $2\sigma$\,C.L., the viable values of $N_{k}$\,{\it e}-folds would satisfy $42\lesssim N_{k}\lesssim 87$ and the rainbow parameter $\lambda$ is nicely constrained to be $\lambda \lesssim6.0$.
   \\[2mm]
\end{abstract}

\maketitle

\section{Introduction} 

In the past three decades, the cosmological inflation model has been well established as the leading paradigm for the very early universe both theoretically and experimentally.  It provides  solutions to important problems of standard big bang cosmology such as the flatness and horizon problem and gives a mechanism for seeds of the formation of the large scale structure of the universe.   Although the inflation model agrees remarkably well with the observational data, it still suffers some fundamental issues, such as the fine-tuning slow-roll potential, the initial conditions, and the trans-Planckian problem.  In addition, the standard mechanism of inflation requires a scalar field (an inflaton) to drive the exponential expansion of the very early universe but physical nature and fundamental origin of an inflaton are still an open question.

Since a period of inflation may occur in the very early universe, one can, in principle, expect that inflation comes from a semi-classical theory of quantum gravity with some quantum correction of Einstein-Hilbert action at high energy.   The simplest example is $f(R)$-gravity theories, where the modification of the action of Einstein-Hilbert action is intended to generalize the Ricci scalar to be some function of the Ricci scalar itself. (for a review see \cite{De_Felice_10} and references therein). In the 4-dimensional action in $f(R)$ gravity, we write the action $S=\frac{1}{2\kappa^{2}}\int d^{4}x\sqrt{-g}f(R)$ where $\kappa^{2} = 8\pi G$, $g$ is the determinant of the metric $g_{\mu\nu}$. The Ricci scalar R is defined by $R = g^{\mu\nu}R_{\mu\nu}$. One can linearize the action by introducing an auxiliary field $\phi$ such that $S=\frac{1}{2\kappa^{2}}\int d^{4}x\sqrt{-g}[F(\phi)(R-\phi) + F(\phi)]$ where we have defined $F(\phi)=dF(\phi)/d\phi$. It is obvious that by making the variation of the action with respect to $\phi$, we simply get $\phi = R$. Thus, a scalar field $\phi$, which will be the inflaton in the scalar field representation, will propagate the freedom degree from modified gravity. In summary, in the case of $f(R)$-gravity for inflation, we can pass to its scalar-field representation to achieve some useful simplification whose mechanism has been well studied in literature, e.g. \cite{Barrow:1988xh,Maeda:1987xf}. The well-known example is the ``Starobinsky model'' for inflation \cite{Starobinsky_80, Mukhanov_81,Starobinsky_83} with a correction quadratic in the Ricci scalar in the modified gravity framework, and of an exponential potential in the scalar field framework.  Starobinsky model offers the clear origin of the inflaton field stemming from gravitational theory by identifying it with the spin-0 part of spacetime metric.  

Adding quantum corrections to the action is not the only possible modification to Einstein's gravity.   Almost all approaches to quantum gravity suggest  that the standard energy momentum dispersion relation is deformed near the Planck scale.  This feature was predicted from string field theory \cite{Kostelecky_89}, loop quantum gravity \cite{Gambini_98}, non-commutative geometry \cite{Carroll_01}.  The modification of the dispersion relation generally takes the form 
\begin{equation}
\varepsilon^2 \tilde{f}^2(\varepsilon) - p^2\tilde{g}^2(\varepsilon) = m^2.
\label{dispersion}
\end{equation}
The functions $\tilde{f}(\varepsilon)$ and $\tilde{g}(\varepsilon)$ are commonly known as the rainbow functions assuming $\tilde{f}(\varepsilon) \rightarrow1$ and $ \tilde{g}(\varepsilon) \rightarrow1$ in the limit $\varepsilon/M \rightarrow 0$ where $M$ is the energy scale that quantum effects of gravity become important. 

One of the interesting approaches that naturally produce modified dispersion relations is called doubly special relativity\cite{Amelino-Camelia_01,Amelino-Camelia_02,Procaccini_05}. In addition to the invariance of the speed of light, it extends special relativity by including an invariant energy scale, usually assumed to be the Planck energy.  Magueijo and Smolin \cite{Magueijo_04} generalized this idea to include curvature i.e. doubly general relativity. In their approach, the spacetime metric felt by a free particle depends on the energy or momentum of the probe particle. Thus, spacetime is represented by a one parameter family of metrics parametrized by energy of the probe $\varepsilon$ , forming a rainbow of metrics, and hence this approach is called gravity's rainbow. The rainbow modified metric can be written as
\begin{equation}
g(\varepsilon) = \eta^{ab} \ \tilde{e}_a(\varepsilon) \otimes \tilde{e}_b(\varepsilon) .
\end{equation}
The energy dependence of the frame field $\tilde{e}_a(\varepsilon)$ can be written in terms of the energy independence frame field $e_a$ as $\tilde{e}_0(\varepsilon) = \frac{1}{\tilde{f}(\varepsilon)} e_0$ and $\tilde{e}_i(\varepsilon) = \frac{1}{\tilde{g}(\varepsilon)} e_i$ for $i = 1,2,3$.  In the study of cosmology, the conventional Friedmann-Robertson-Walker (FRW) metric for the homogeneous and isotropic universe is replaced by a rainbow metric of the form
\begin{equation}
ds^2 = -\frac{1}{\tilde{f}^2(\varepsilon)} dt^2 + a^2(t)\delta_{ij}dx^idx^j.
\label{FRW}
\end{equation}
For simplicity, we chose $\tilde{g}(\varepsilon) = 1$ and only considered the spatially flat case with $K = 0$.  In general, the rainbow function $\tilde{f}(\varepsilon)$ does not depend on space-time coordinates since for any specific operation of measurement the probe's energy $\varepsilon$ can be treated as a constant.  However, instead of considering any specific measurement, the author in \cite{Ling07} suggested that this formalism  can be generalized to study semi-classical effects of relativistic particles on the background metric during a longtime process.  For the very early universe, we can choose massless particles which dominate the universe at that period such as gravitons and inflatons as our probes.   In this case, we need to consider the evolution of the probe's energy with cosmic time, denoted as $\varepsilon(t)$.  As a result the rainbow functions $\tilde{f}(\varepsilon)$ depends on time only implicitly through the energy of particles.  

The rainbow universe formalism was studied in the case of $f(R) = R$ \cite{Ling07,Ling08} by using modified Friedmann equations and the early universe is driven by thermally fluid substance.   The cosmological linear perturbations of this model was studied in \cite{Wang14}.  In this paper, we extend this formalism to more general $f(R)$ gravity, in particular the Starobinsky model.   Qualitatively, Eq. (\ref{FRW}) implies that time and space are scaled with different scaling functions i.e. the rainbow function and the scale factor respectively.  By choosing suitable form of $\tilde{f}(\varepsilon)$, one can solve the horizon problem without any need of inflationary expansion \cite{Moffat_93,Albrecht_99,Barrow_99}.  In our case, however, the quadratic term in the Ricci scalar will lead to an inflationary solution for $a(t)$.  We will show how the observational data put the constraint on the scaling function $\tilde{f}(\varepsilon)$. 

We organize the paper as follows.  In section \ref{equation}, we derive the equations of motion for $f(R)$ theory in the framework of rainbow universe where cosmological evolution of probes is taken into account and written down as an inflationary solution to these equations for the case of Starobinsky model.  Then we turn to investigate cosmological perturbation in the rainbow universe in section \ref{perturbation} where the spectral index of scalar perturbation and the tensor-to-scalar ratio of the model will be calculated.  In section \ref{data}, we use the Planck 2015 data set to constraint the rainbow parameter of our model.  Finally, we present our conclusion in section \ref{con}.

%%%%%%%%%%%%%%%%%%%%%%%%%%%%%%%%%%%%%%%%%%%%%%%%%%%%%%%%%%%%%%%%%%%%%%%%%%%%%%%%%%%%%%%%%%%%%%%%%%%%%%%%%%%%%%%%%%%%%%%%%%%%%%%%%%%%%%%%%
\section{$f(R)$ theory with Rainbow gravity's effect}\label{equation}
 
In addition to the modified FRW metric (\ref{FRW}), modifications to General Relativity are expected to be possible in the very early universe where some corrections to Einstein's gravity may emerge at high curvature.  The simplest class of modified gravity theory is $f(R)$ gravity, where the Einstein-Hilbert term in the action is replaced by a function of the Ricci scalar.  Now taking this into account we derive the field equations by following the formalism in $f(R)$ theory described in \cite{De_Felice_10}.   We start by considering the 4-dimensional action in $f(R)$ gravity:
\begin{equation}
S = \frac{1}{2\kappa^2} \int d^4x\sqrt{-g} f(R) + \int d^4x \sqrt{-g}\mathcal{L}_M,
\label{Action}
\end{equation}
where $\kappa^2 = 8\pi G$ and $\mathcal{L}_M$ is the matter Lagrangian.  By varying the action (\ref{Action}) with respect to $g_{\mu\nu}$, we obtain the field equations:
\begin{widetext}
\begin{align}
\Sigma_{\mu\nu} \equiv F(R)R_{\mu\nu}(g) - \frac{1}{2}f(R)g_{\mu\nu}-\nabla_\mu\nabla_\nu F(R)+g_{\mu\nu} \square F(R) = \kappa^2 T_{\mu\nu}^{(M)},
\label{eom}
\end{align}
\end{widetext} 
where $F(R) \equiv \partial f(R)/\partial R$ and $\square F = (1/\sqrt{-g})\partial_\mu (\sqrt{-g}g^{\mu\nu}\partial_\nu F)$. The energy-momentum tensor of matter fields is defined by $T_{\mu\nu}^{(M)} \equiv - \frac{2}{\sqrt{-g}}\frac{\delta(\sqrt{-g} \mathcal{L}_M)}{\delta g^{\mu\nu}}$.   $\Sigma_{\mu\nu}$ and $ T^{(M)}_{\mu\nu}$ both satisfy the continuity equation i.e. $\nabla^\mu \Sigma_{\mu\nu} = 0$ and $\nabla^\mu T^{(M)}_{\mu\nu} =0 $ respectively.   In cosmology, the energy-momentum tensor of matter is given in the perfect fluid form as  $T^{\mu (M)}_{\nu} = \text{diag}(-\rho_M,P_M,P_M,P_M)$, where $\rho_M$ is the energy density and $P_M$ is the pressure.  In the following, we will derive cosmological solutions to field equations (\ref{eom}).  In so doing, let us assume the modified FRW background (\ref{FRW}), we obtain
\begin{align}
3(F H^2 + H \dot{F}) + \dot{F}\frac{\dot{\tilde{f}}}{\tilde{f}} =  \frac{FR-f(R)}{2\tilde{f}^2} + \frac{\kappa^2 \rho_M}{\tilde{f}^2},
\label{fR_eom_1}
\end{align}
\begin{align}
\ddot{F} - H\dot{F} + 2 F \dot{H} + 2 FH\frac{\dot{\tilde{f}}}{\tilde{f}} =  -\frac{\kappa^2}{\tilde{f}^2}(\rho_M+P_M),
\label{fR_eom_2}
\end{align}
where the Hubble parameter $H = \dot{a}/a$ and $\dot{a} = da/dt$.  

%%%%%%%%%%%%%%%%%%%%%%%%%%%%%%%%%%%%%%%%%%%%%%%%%%%%%%%%%%%%%%%%%%%%%%%%%%%%%%%%%%%%%%%%%%%%%%%%%%%%%%%%%%%%%%%%%%%%%%%%%%%%%%%%%%%%%%%%%
\subsection{Inflation with the effect of Rainbow Gravity}\label{Inflation}

In the following let us focus on the Starobinsky's model \cite{Starobinsky_80} in which $f(R)$ takes the following form
\begin{equation}
f(R) = R + R^2/(6M^2),
\end{equation} 
where $M$ has a dimension of mass and is chosen to be the same as the mass scale for quantum gravity in Eq. (\ref{dispersion}) for simplicity.   This is the first inflation model related to the conformal anomaly in quantum gravity.   The $R^2$ term is responsible for the de Sitter expansion.  Because of the linear term in $R$, the inflationary expansion ends when the term $R^2/(6M^2)$ becomes smaller than the linear term $R$. This gives the graceful exit to inflation.

In contrast to previous studies on rainbow universe \cite{Ling07,Ling08,Wang14}, the $R^2$-term plays a leading role in the very early universe instead of relativistic matter.  We will set $\rho_M = 0$ and $P_M =0$ from now on. One can also consider this assumption in a scalar field framework of $f(R)$ theory.  For the case of Starobinsky model, its classical dynamics would be equivalent to the system of one scalar (an inflaton) with an exponential potential.  So, it is reasonable to assume that inflatons dominated the very early universe and neglect contributions from matter and radiation.  Then the combination of equation (\ref{fR_eom_1}) and (\ref{fR_eom_2}) gives us
\begin{widetext}
\begin{align}
\ddot{H}-\frac{\dot{H}^2}{2H} + \frac{1}{2\tilde{f}^2}M^2H + \frac{11}{6} H\left( \frac{\dot{\tilde{f}}}{\tilde{f}}\right)^2  + \frac{1}{3}\left(\frac{\dot{\tilde{f}}}{ \tilde{f}}\right)^3 +\frac{10}{3}\dot{H}\frac{\dot{\tilde{f}}}{\tilde{f}}+\frac{\dot{H}}{H}\left( \frac{\dot{\tilde{f}}}{\tilde{f}}\right)^2+H\frac{\ddot{\tilde{f}}}{\tilde{f}} + \frac{1}{3}\frac{\dot{\tilde{f}}\ddot{\tilde{f}}}{\tilde{f}^2}+\frac{1}{3}\frac{\ddot{H}}{H}\frac{\dot{\tilde{f}}}{\tilde{f}} = -3H^2\left(\frac{\dot{\tilde{f}}}{\tilde{f}} +\frac{\dot{H}}{H}\right)\,.\label{R2_01}
\end{align}
\end{widetext} 
 and
 \begin{equation}
 \ddot{R} + 3H\dot{R}+\frac{4\dot{\tilde{f}}\dot{R}}{3\tilde{f}} + \frac{M^2R}{\tilde{f}^2} = 0.
 \label{R2_02}
 \end{equation}
Now, we consider the evolution of modified FRW universe.  For explicitness, we need to choose the form of $\tilde{f}(\varepsilon)$ and this can be explained in the ``Einstein frame," where the scalar degree of freedom takes a canonical form with a potential.  Under the conformal transformation $(g_E)_{\mu\nu} = \Omega^2 g_{\mu\nu}$ with $\Omega^2 = F$, the action in Einstein frame can be written as 
\begin{equation}
S_E = \int d^4x\sqrt{-g_E} \left[\frac{1}{2\kappa^2} R_E -\frac{1}{2}g_E^{\mu\nu} \partial_\mu \phi_E \partial_\nu \phi_E - V(\phi_E) \right],
\end{equation}
where the subscribe $E$ denote quantities in Einstein frame.  We define the scalar field (or the inflaton) by $\kappa\phi_E \equiv \sqrt{3/2}\ln F$ and its potential is determined by $V(\phi_E) = (FR-f)/(2\kappa^2F^2)$.  Note that we neglect contribution from other matters. 

In this framework it is natural to choose an inflaton as our probe during the inflation.  We assume that probe's energy $\varepsilon_E(t)$ is proportional to the total energy density of the inflaton, $\rho_{\phi_E}$.   During the slow-roll inflation, we can assume further that $\rho_{\phi_E} \approx V(\phi_E)$ and it can be treated as a function of the Ricci Scalar in the Jordan frame, $R$.  In the case of FRW metric ($\tilde{f} = 1$), for the Starobinsky model, it is well known that $R \approx 12H^2 - M^2$ during inflation \cite{De_Felice_10}.  Hence, we can now assume that the probe energy $\varepsilon_E(t)$ can be expressed in terms of a geometric expansion rate $H$ and this conclusion should also be applied in the Jordan frame.
%The field equation (\ref{eom}) can be written in the form of Einstein equation $G_{\mu\nu} = \kappa^2 T^{(D)}_{\mu\nu}$ where $G_{\mu\nu} \equiv R_{\mu\nu} - (1/2) g_{\mu\nu} R$ and 
%\begin{equation}
%\kappa^2 T^{(D)}_{\mu\nu} \equiv g_{\mu\nu}(f-R)/2 + \nabla_{\mu}\nabla_{\nu}F - g_{\mu\nu} \square F - (1-F)R_{\mu\nu}.\nonumber
%\end{equation}
%We can see that $T^{(D)}_{\mu\nu}$ plays the role of effective energy-momentum tensor.  By assuming unmodified FRW metric, the 00 components of Einstein equation leads us the Friedmann equation $H^2 = \kappa^2 \rho^{(D)}$ where $\rho^{(D)} \equiv T^{(D)}_{00}$ represents the effective energy density of the universe.  We then assume that the energy $\varepsilon(t)$ is proportional to the total energy density $\rho^{(D)}$ as in \cite{Ling07} and can be related to the Hubble parameter $H$ via Friedman equation.  

Based on these assumptions, we choose the rainbow function to be explicitly parametrized in the power-law form of the Hubble parameter i.e. $\tilde{f}^2 = 1 + (H/M)^{2\lambda}$ with the rainbow parameter $\lambda > 0$ and $\tilde{f} =1$ for late time.  For the inflationary period, we are interested in the limit $H^2 \gg M^2$. Then, the rainbow function can be approximated as $\tilde{f} \approx (H/M)^\lambda$.  Equation (\ref{R2_01}) can be written as
\begin{widetext}
\begin{align}
\frac{1}{2(1+\lambda)}\frac{M^2H}{\tilde{f}^2} + 3H\dot{H} + \frac{1}{6}(17\lambda-3)\frac{\dot{H}^2}{H} + \frac{2\lambda^2\dot{H}^3}{3H^3} + (1+\frac{\lambda}{3}\frac{\dot{H}}{H^2})\ddot{H} =0.
\label{R2_03}
\end{align}
\end{widetext} 
During inflation $H$ is very slowly varying.  We can keep the first two terms in Equation (\ref{R2_03}) and neglect others.  This gives us
\begin{equation}
\dot{H} \simeq - \frac{M^{2\lambda+2}}{6(1+\lambda)}H^{-2\lambda}.
\end{equation}
We then obtain the solution
\begin{align}
&H \simeq H_i - \frac{M^2}{6(1+\lambda)}\left(\frac{M}{H_i}\right)^{2\lambda}(t-t_i),\label{H_evolution}\\&
a \simeq a_i \exp\left\{H_i(t-t_i) - \frac{M^2}{12(1+\lambda)}\left(\frac{M}{H_i} \right)^{2\lambda}(t-t_i)^2 \right\},\\&
R \simeq 12H^2 \left(\frac{H}{M}\right)^{2\lambda},
\end{align}
where $H_i$ and $a_i$ are Hubble parameter and scale factor at the onset of inflation ($t = t_i$), respectively.  The slow-roll parameter is defined by
\begin{equation}
\epsilon_1 = -\frac{\dot{H}}{H^2} = \frac{1}{6(1+\lambda)}\left(\frac{M}{H} \right)^{2+2\lambda},
\end{equation}
which is less than unity in the limit $H^2 \gg M^2$.  One can also check that the approximate relation $3H\dot{R} +\frac{M^2R}{\tilde{f}^2} \simeq 0$ holds in Eq. (\ref{R2_02}) by using $R \simeq 12H^2 \left(\frac{H}{M}\right)^{2\lambda}$.  The end of inflation (at time $t = t_f$) can be determined by the condition $\epsilon_1 \simeq 1$, i.e. $H_f \simeq M/[6(1+\lambda)]^{1/(2+2\lambda)}$.  The number of $e$-foldings from $t = t_i$ to $t = t_f$ is defined by
\begin{align}
N &\equiv \int_{t_i}^{t_f} Hdt \nonumber\\&\simeq H_i(t_f-t_i) - \frac{M^2}{12(1+\lambda)}\left(\frac{M}{H_i} \right)^{2\lambda}(t_f-t_i)^2.
\end{align}
Since inflation ends at $t_f \simeq t_i + 6(1+\lambda) H_i^{1+2\lambda}/M^{2+2\lambda}$, we can show that
\begin{equation}
N \simeq 3(1+\lambda)\left(\frac{H_i}{M} \right)^{2+2\lambda} \simeq \frac{1}{2\epsilon_1(t_i)}. \label{N_fold}
\end{equation}
Note that for $\lambda = 0$ the model reverts to the Starobinsky model. 

%%%%%%%%%%%%%%%%%%%%%%%%%%%%%%%%%%%%%%%%%%%%%%%%%%%%%%%%%%%%%%%%%%%%%%%%%%%%%%%%%%%%%%%%%%%%%%%%%%%%%%%%%%%%%%%%%%%%%%%%%%%%%%%%%%%%%%%%%
\section{Cosmological perturbation in Rainbow gravity}\label{perturbation}
In this section, we focus on cosmological linear perturbation generated during inflation by following the formalism presented in \cite{De_Felice_10}.   From the point of view of a probe particle with energy $\varepsilon$, a general perturbed metric about the flat FRW universe would take the form
\begin{widetext}
\begin{align}
 ds^2 = -\frac{1+2\alpha}{\tilde{f}^2(\varepsilon)}dt^2 - \frac{2a(t)(\partial_i\beta-S_i)}{\tilde{f}(\varepsilon)}dtdx^i + a^2(t)(\delta_{ij}+2\psi\delta_{ij}+2\partial_i\partial_j\gamma+2\partial_jF_i+h_{ij})dx^idx^j\,, \label{B1}
\end{align}
\end{widetext} 
where $\alpha$, $\beta$, $\psi$, $\gamma$ are scalar perturbations, $S_i$, $F_i$ are vector perturbations, and $h_{ij}$ are tensor perturbations.  The rainbow function $\tilde{f}(\varepsilon)$ can be viewed as a smooth background function.   As usual, we will focus on scalar and tensor perturbations by setting vector perturbations $S_i=F_i=0$.  Before discussing the evolution of cosmological perturbations, we construct a number of gauge-invariant quantities required to eliminate unphysical modes. Let us consider the gauge transformation $t \rightarrow \hat{t} = t+\delta t$ and $x^i \rightarrow \hat{x}^i = x^i + \delta^{ij}\partial_j \delta x$.  Here $\delta t$ and $\delta x$ denote the infinitesimal scalar functions of the spacetime coordinates.  Then we can show that the scalar metric perturbations transform as: 
\begin{eqnarray}
\alpha &\rightarrow& \hat{\alpha} = \alpha +\frac{\dot{\tilde{f}}}{\tilde{f}}\delta t - \dot{\delta t}, \\
\beta &\rightarrow& \hat{\beta} = \beta - \frac{\delta t}{a\tilde{f}} + a \tilde{f} \dot{\delta x},\\
\psi &\rightarrow& \hat{\psi} = \psi - H\delta t,\\
\gamma &\rightarrow& \hat{\gamma} = \gamma - \delta x.
\end{eqnarray}
Note that the tensor perturbations $h_{ij}$ are invariant under the gauge transformation.   We can define gauge invariant quantities under the above gauge transformations as
\begin{eqnarray}
\Phi &=& \alpha - \tilde{f}\frac{d}{dt}\left[ a^2\tilde{f}\left(\dot{\gamma} + \frac{\beta}{a\tilde{f}} \right)\right],\\
\Psi &=& -\psi + a^2\tilde{f}^2H\left(\dot{\gamma}+\frac{\beta}{a\tilde{f}} \right),\\
\mathcal{R} &=& \psi - \frac{H\delta F}{\dot{F}}.
\end{eqnarray}
We can choose the longitudinal gauge $\beta = 0$ and $\gamma=0$.    In this gauge $\Phi = \alpha$ and $\Psi = -\psi$.  Thus, without tensor perturbations, the line element takes the form
\begin{eqnarray}
 ds^2 = -\frac{1+2\Phi}{\tilde{f}^2(t)}dt^2 + a^2(t)(1-2\Psi)\delta_{ij}dx^idx^j .\label{B2}
\end{eqnarray}
For later convenience, we define the perturbed quantity
\begin{eqnarray}
 A \equiv 3(H\Phi+\dot{\Psi})  . \label{B3}
\end{eqnarray}
By using the perturbed metric (\ref{B2}) and Eq. (\ref{eom}), we come up with the following equations
\begin{widetext}
\begin{align}
\frac{\nabla^2\Psi}{a^2}+\tilde{f}^2HA = -\frac{1}{2F}\left[3\tilde{f}^2\left(H^2 + \dot{H} + \frac{\dot{\tilde{f}}}{\tilde{f}}\right)\delta F + \frac{\nabla^2\delta F}{a^2}-3\tilde{f}^2H\delta\dot{F} + 3\tilde{f}^2H\dot{F}\Phi + \tilde{f}^2\dot{F}A+\kappa^2\delta\rho_M\right]\ , \label{B4}
\end{align}
\end{widetext} 
\begin{eqnarray}
H\Phi+\dot{\Psi}=-\frac{1}{2F}(H\delta F+\dot{F}\Phi-\delta\dot{F}) \ , \label{B5} 
\end{eqnarray}
and
\begin{widetext}
\begin{align}
\dot{A} + \left(2H+\frac{\dot{\tilde{f}}}{\tilde{f}}\right)A+3\dot{H}\Phi + \frac{\nabla^2\Phi}{a^2\tilde{f}^2}+\frac{3H\Phi\dot{\tilde{f}}}{\tilde{f}}\nonumber\\= \frac{1}{2F}\left[3\delta\ddot{F}+3\left(H+\frac{\dot{\tilde{f}}}{\tilde{f}}\right)\delta\dot{F}-6H^2\delta F -\frac{\nabla^2\delta F}{a^2\tilde{f}^2} - 3\dot{F}\dot{\Phi}-\dot{F}A - 3\left(H+\frac{\dot{\tilde{f}}}{\tilde{f}}\right)\dot{F}\Phi-6\ddot{F}\Phi+\frac{\kappa^2}{\tilde{f}^2}(3\delta P_M+\delta\rho_M) \right] .
\label{B6}
\end{align}
\end{widetext} 
These equations describe evolutions of the scalar perturbations.  We shall solve the above equations  in the context of 
the inflationary universe. 
%%%%%%%%%%%%%%%%%%%%%%%%%%%%%%%%%%%%%%%%%%%%%%%%%%%%%%%%%%%%%%%%%%%%%%%%%%%%%%%%%%%%%%%%%%%%%%%%%%%%%%%%%%%%%%%%%%%%%%%%%%%%%%%%%%%%%%%%%%%%%%
\subsection{Curvature perturbations}
Let us consider scalar perturbations generated during inflation without taking into account the perfect fluid i.e. $\delta\rho_M =0$ and $\delta P_M=0$.  We can choose the gauge condition $\delta F =0$, so that $\mathcal{R} = \psi = -\Psi$.  Since the spatial curvature $^{(3)}\mathcal{R}$ on the constant-time hypersurface is related to $\psi$ via the relation $^{(3)}\mathcal{R} = - 4\nabla^2\psi/a^2$, the quantity $\mathcal{R}$ is often called the curvature perturbation on the uniform-field hypersurface.  By setting $\delta F$ to be zero, Eq. (\ref{B5}) gives us 
\begin{eqnarray}
 \Phi = \frac{\dot{\mathcal{R}}}{H+\dot{F}/2F} \ , \label{B8}
\end{eqnarray}
and from the equation (\ref{B5}), we get
\begin{eqnarray}
 A = -\frac{1}{H+\dot{F}/2F}\left[\frac{\nabla^2\mathcal{R}}{a^2\tilde{f}^2}+\frac{3H\dot{F}\dot{\mathcal{R}}}{2F(H+\dot{F}/2F)}\right] \ . \label{B9}
\end{eqnarray}
By using the background equation (\ref{fR_eom_2}), Eq. (\ref{B4}) gives
\begin{widetext}
\begin{align}
 \dot{A}+\left(2H+\frac{\dot{F}}{2F}\right)A+\frac{\dot{\tilde{f}}A}{\tilde{f}}+\frac{3\dot{F}\dot{\Phi}}{2F}+\left[\frac{3\ddot{F}+6H\dot{F}}{2F}+\frac{\nabla^2}{a^2\tilde{f}^2}\right]\Phi+\frac{3\dot{F}}{2F}\frac{\Phi\dot{\tilde{f}}}{\tilde{f}} = 0. \label{B10}
\end{align}
\end{widetext} 
Substituting Eq. (\ref{B8}) and (\ref{B9}) into Eq. (\ref{B10}), we can show that the curvature perturbation satisfies the following equation in Fourier space
\begin{eqnarray}
 \ddot{\mathcal{R}} + \frac{1}{a^3Q_s}\frac{d}{dt}(a^3Q_s)\dot{\mathcal{R}} + \frac{\dot{\tilde{f}}}{\tilde{f}}\dot{\mathcal{R}} + \frac{k^2}{a^2\tilde{f}^2} \mathcal{R}= 0 \ , \label{B13}
\end{eqnarray}
where $k$ is a comoving wave number and $Q_s$ is defined by
\begin{eqnarray}
 Q_s \equiv \frac{3\dot{F}^2}{2\kappa^2F(H+\dot{F}/2F)^2} \ . \label{B12}
\end{eqnarray}
To simplify Eq. (\ref{B13}) any further we will introduce the new variables $z_s = a\sqrt{Q_s}$ and $u = z_s\mathcal{R}$.  The above equation can be expressed as   
\begin{eqnarray}
 u'' + \left(k^2-\frac{z_s''}{z_s}\right)u = 0 \ , \label{B14}
\end{eqnarray}
where a prime denotes a derivative with respect to the new time coordinates $\eta = \int (a\tilde{f})^{-1} dt$.  To calculate the spectrum of curvature perturbations, let us further introduce the Hubble flow parameters (also known as slow-roll parameters)  as
\begin{eqnarray}
 \epsilon_1 \equiv -\frac{\dot{H}}{H^2},  \ \ \epsilon_3 \equiv \frac{\dot{F}}{2HF}, \ \ \epsilon_4 \equiv \frac{\dot{E}}{2HE}\ ,
\end{eqnarray}
where $E \equiv 3\dot{F}^2/2\kappa^2$.   Using these definitions, $Q_s$ can be rewritten as
\begin{eqnarray}
 Q_s = \frac{E}{FH^2(1+\epsilon_3)^2} \ . \label{B16}
\end{eqnarray}
Assuming that parameters $\epsilon_i$ are nearly constants during the inflation for $i=1,3,4$ and $\tilde{f} \simeq (H/M)^{\lambda}$, we have $\eta = -1/[(1-(1+\lambda)\epsilon_1)\tilde{f}aH]$.   If $\dot{\epsilon_i}\simeq 0$, it follows that the term $z_s''/z_s$ is given by
\begin{eqnarray}
 \frac{z_s''}{z_s} = \frac{\nu^2_{\mathcal{R}} - 1/4}{\eta^2} \ , \label{B18}
\end{eqnarray}
with
\begin{eqnarray}
 \nu_{\mathcal{R}}^2 = \frac{1}{4} + \frac{(1+\epsilon_1 - \epsilon_3+\epsilon_4)(2-\lambda\epsilon_1 -\epsilon_3+\epsilon_4)}{(1-(\lambda+1)\epsilon_1)^2} \ .\label{B19}
\end{eqnarray}
With these assumptions, the explicit solution for Eq. (\ref{B14}) can be found as
\begin{eqnarray}
 u = \frac{\sqrt{\pi|\eta|}}{2}\textmd{e}^{i(1+2\nu_{\mathcal{R}})\pi/4}\left[c_1\textmd{H}_{\nu_{\mathcal{R}}}^{(1)}(k|\eta|)+c_2\textmd{H}_{\nu_{\mathcal{R}}}^{(2)}(k|\eta|)\right] \ , \label{B20}
\end{eqnarray}
where $c_1$, $c_2$ are integration constants and $\textmd{H}_{\nu_{\mathcal{R}}}^{(1)}(k|\eta|)$, $\textmd{H}_{\nu_{\mathcal{R}}}^{(2)}(k|\eta|)$ are the Hankel functions of the first kind and the second kind respectively.
In the asymptotic past $k\eta \rightarrow -\infty$, the solution to Eq. (\ref{B14}) is  $u \rightarrow \textmd{e}^{-ik\eta}/\sqrt{2k}$, this implies $c_1=1$ and $c_2=0$.  Thus we have
\begin{eqnarray}
 u = \frac{\sqrt{\pi|\eta|}}{2}\textmd{e}^{i(1+2\nu_{\mathcal{R}})\pi/4}\textmd{H}_{\nu_{\mathcal{R}}}^{(1)}(k|\eta|) \ . \label{B21}
\end{eqnarray}
The power spectrum of curvature perturbations is defined by
\begin{eqnarray}
 \mathcal{P}_{\mathcal{R}} \equiv \frac{4\pi k^3}{(2\pi)^3}|\mathcal{R}|^2 \ . \label{B22}
\end{eqnarray}
By using Eq. (\ref{B21}) and the relation $u = z_s\mathcal{R}$, we will get
\begin{eqnarray}
 \mathcal{P}_{\mathcal{R}} = \frac{1}{Q_s}\left[(1-(1+\lambda)\epsilon_1)\frac{\Gamma(\nu_{\mathcal{R}})H}{2\pi\Gamma(3/2)}\left(\frac{H}{M}\right)^\lambda\right]^2\left(\frac{k|\eta|}{2}\right)^{3-2\nu_{\mathcal{R}}} \ . \label{B23}
\end{eqnarray}
Note that for $k|\eta| \rightarrow 0$ we have $\textmd{H}_{\nu_{\mathcal{R}}}^{(1)}(k|\eta|) \rightarrow -(i/\pi)\Gamma(\nu_{\mathcal{R}})(k|\eta|/2)^{-\nu_{\mathcal{R}}}$ and $P_{\mathcal{R}}$ should be evaluated at $k=aH$ because $\mathcal{R}$ is fixed after the Hubble radius crossing. Now, we can define the spectral index $n_{\mathcal{R}}$  as
\begin{eqnarray}
 n_{\mathcal{R}} - 1 = \left.\frac{d\textmd{ln}\mathcal{P}_{\mathcal{R}}}{d\textmd{ln}k}\right|_{k=aH} = 3 - 2\nu_{\mathcal{R}} \ . \label{B25}
\end{eqnarray}
During the inflationary epoch, we assume that $|\epsilon_i | \ll 1$ for all $i$, then the spectral index is reduced to
\begin{eqnarray}
 n_{\mathcal{R}} - 1 \simeq -2(\lambda+2)\epsilon_1+2\epsilon_3-2\epsilon_4 \ . \label{B27}
\end{eqnarray}
Giving that $|\epsilon_i|$ are much smaller than one, $n_{\mathcal{R}} \simeq 1$, the spectrum is nearly scale-invariant.   Subsequently, the power spectrum of curvature perturbation is
\begin{eqnarray}
 \mathcal{P}_{\mathcal{R}} \approx \frac{1}{Q_s}\left(\frac{H}{2\pi}\right)^2\left(\frac{H}{M}\right)^{2\lambda} \ . \label{B28}
\end{eqnarray}

%%%%%%%%%%%%%%%%%%%%%%%%%%%%%%%%%%%%%%%%%%%%%%%%%%%%%%%%%%%%%%%%%%%%%%%%%%%%%%%%%%%%%%%%%%%%%%%%%%%%%%%%%%%%%%%%%%%%%%%%%%%%%%%%%%%%%%%%%%%%%
\subsection{Tensor perturbations}
In this section we derive appropriate expressions of the power spectrum and the spectral index of tensor perturbations.   In general,  the tensor perturbation $h_{ij}$ can be generally written as
\begin{eqnarray}
 h_{ij} = h_{+}e^+_{ij} + h_{\times}e^\times_{ij} \ , \label{C2}
\end{eqnarray}
where $e^+_{ij}$ and $e^\times_{ij}$ are the polarization tensors corresponding to the two polarization states of $h_{ij}$.  Let $\vec{k}$ be in the direction along the z-axis, then the non-vanishing components of polarization tensors are $e^+_{xx} = -e^+_{yy} = 1$ and $e^\times_{xy} = e^\times_{yx} = 1$.   By neglecting the scalar and vector perturbation, the perturbed FRW metric in (\ref{B1}) can be written as
\begin{widetext}
\begin{align}
 ds^2 = -\frac{dt^2}{\tilde{f}(\varepsilon)^2} + a^2(t)h_{\times}dxdy + a^2(t)\left[(1+h_{+})dx^2+(1-h_{+})dy^2+dz^2\right] .\label{C1}
\end{align}
\end{widetext} 
By using the equation of motion (\ref{eom}), we can show that the Fourier components $h_\chi$  satisfy the equation
\begin{eqnarray}
  \ddot{h}_\chi + \frac{(a^3F)^\cdot}{a^3F}\dot{h}_\chi + \frac{\dot{\tilde{f}}}{\tilde{f}}\dot{h}_\chi + \frac{k^2}{a^2\tilde{f}^2}h_\chi = 0 \ , \label{C5}
\end{eqnarray}
where $\chi$ denotes $+$ and $\times$.  We now follow the similar procedure to the one given in the case of curvature perturbation.   Let us introduce the new variables $z_t = a\sqrt{F}$ and $u_\chi = z_t h_\chi /\sqrt{16\pi G}$, then Eq. (\ref{C5}) can be written as
\begin{eqnarray}
 u''_\chi + \left(k^2-\frac{z_t''}{z_t}\right)u_\chi = 0 \ . \label{C6}
\end{eqnarray}
Note that the massless scalar field $u_\chi$ has dimension of mass.   By choosing $\dot{\epsilon}_i=0$, we get
\begin{eqnarray}
 \frac{z_t''}{z_t} = \frac{\nu^2_t-1/4}{\eta^2} \ , \label{C7}
\end{eqnarray}
with
\begin{eqnarray}
 \nu^2_t = \frac{1}{4} + \frac{(1+\epsilon_3)(2-(1+\lambda)\epsilon_1+\epsilon_3)}{(1-(1+\lambda)\epsilon_1)^2} \ . \label{C8}
\end{eqnarray}
Hence, the solution to the equation (\ref{C6}) can be expressed in terms of  the Hankel functions as in Eq. (\ref{B21}).  Taking into account polarization
states, the power spectrum of tensor perturbations $P_T$ after the Hubble radius crossing is  
\begin{align}
 \mathcal{P}_T &= 4\times\frac{16\pi G}{a^2F}\frac{4\pi k^3}{(2\pi)^3}|u_\chi|^2  \nonumber\\&= \frac{16}{\pi}\left(\frac{H}{m_{pl}}\right)^2\frac{1}{F}\left[(1-(1+\lambda)\epsilon_1)\frac{\Gamma(\nu_t)}{\Gamma(3/2)}\left(\frac{H}{M}\right)^\lambda\right]^2\times\nonumber\\&\,\,\,\,\,\times\left(\frac{k|\eta|}{2} \right)^{3-2\nu_t}.  
 \label{C12}
\end{align}
Note that we have used $\tilde{f} \simeq (H/M)^{\lambda}$.  Since all of the slow-roll parameters are very small during the inflation ($|\epsilon_i| \ll 1$), $\nu_t$ can be estimated as
\begin{eqnarray}
 \nu_t \simeq \frac{3}{2} + (1+\lambda)\epsilon_1 + \epsilon_3 \ . \label{C13}
\end{eqnarray}
In addition, the spectral index of tensor perturbations is determined by
\begin{eqnarray}
 n_T = \left.\frac{d \textmd{ln}\mathcal{P}_T}{d\textmd{ln}k}\right|_{k=aH} = 3-2\nu_t \simeq -2(1+\lambda)\epsilon_1 - 2\epsilon_3 \ , \label{C16}
\end{eqnarray}
the power spectrum $\mathcal{P}_T$ can also be rewritten as
\begin{eqnarray}
 \mathcal{P}_T \simeq \frac{16}{\pi}\left(\frac{H}{m_{pl}}\right)^2\frac{1}{F}\left(\frac{H}{M}\right)^{2\lambda} \ . \label{C17}
\end{eqnarray}
The tensor-to-scalar ratio $r$ can be obtained as
\begin{eqnarray}
 r \equiv \frac{\mathcal{P}_T}{\mathcal{P}_R}  \simeq \frac{64\pi}{m_{pl}^2}\frac{Q_s}{F} \ . \label{C20}
\end{eqnarray}
By substituting $Q_s$ from Eq. (\ref{B16}), we therefore get
\begin{eqnarray}
 r = 48\epsilon_3^2 \ . \label{C21}
\end{eqnarray}
%%%%%%%%%%%%%%%%%%%%%%%%%%%%%%%%%%%%%%%%%%%%%%%%%%%%%%%%%%%%%%%%%%%%%%%%%%%%%%%%%%%%%%%%%%%%%%%%%%%%%%%%%%%%%%%%%%%%%%%%%%%%%%%%%%%%%%%%%
\subsection{The Spectra of Perturbations based on Starobinsky's model in Gravity's rainbow theory}
Let us start by finding relations among the Hubble flow parameters.  In the absence of matter field with an assumption that $ | \epsilon_i | \ll 1$ during the inflation, Eq. (\ref{fR_eom_2}) gives us
\begin{eqnarray}
 \epsilon_3 \simeq -(1+\lambda)\epsilon_1 \ . \label{C23}
\end{eqnarray}
For Starobinsky model, $ f(R) = R + R^2/6M^2$, inflation occurred in the limit $R \gg M^2$ and $|\dot{H}| \ll H^2$.   We can approximate $F(R) \simeq \frac{4H^2\tilde{f}^2}{M^2}$.  By assuming that $|\epsilon_i| \ll 1$ during the inflation, this leads to
\begin{eqnarray}
 \epsilon_4 \simeq -(1+2\lambda)\epsilon_1 \ . \label{C26}
\end{eqnarray}
The spectral index of scalar perturbations and the tensor-to-scalar ratio can be rewritten in terms of $\epsilon_1$ as
\begin{eqnarray}
 n_\mathcal{R} - 1 \simeq -4\epsilon_1\  \  \text{and    } \ r \simeq 48(\lambda+1)^2\epsilon_1^2 . \label{C28} 
\end{eqnarray}
Let $t_k$ be the time at the Hubble radius crossing ($k=aH$).  From Eq. (\ref{H_evolution}), as long as the condition $\frac{M^2(t_k-t_i)}{6(1+\lambda)}(M/H_i)^{2\lambda} \ll H_i$ is satisfied, we can approximate $H(t_k) \simeq H_i$.  The number of $e$-fold from $t=t_k$ to the end of the inflation can be estimated as $N_k \simeq 1/2\epsilon_1(t_k)$.
\begin{figure*}
\begin{center}
\includegraphics[width=0.42\linewidth]{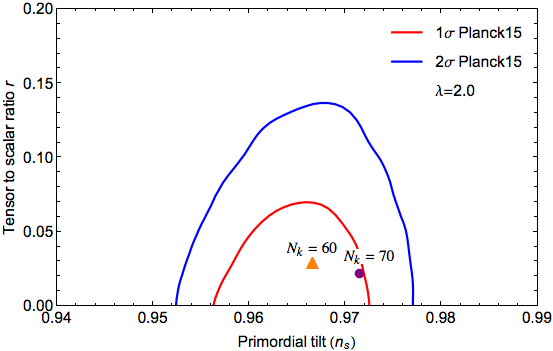}
\includegraphics[width=0.42\linewidth]{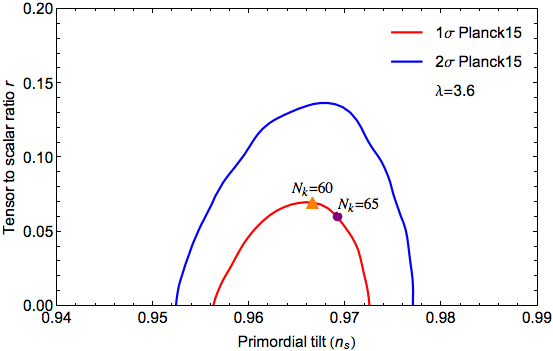}
\includegraphics[width=0.42\linewidth]{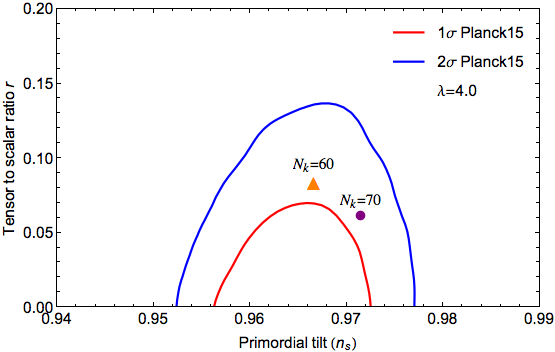}
\includegraphics[width=0.42\linewidth]{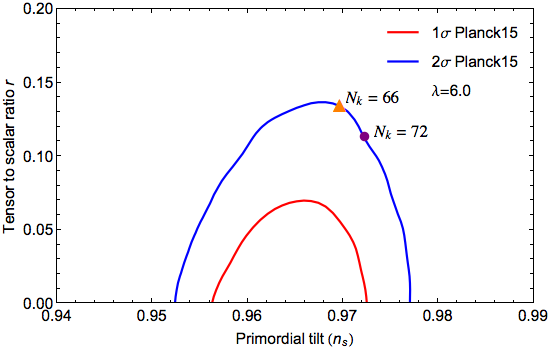}
\includegraphics[width=0.42\linewidth]{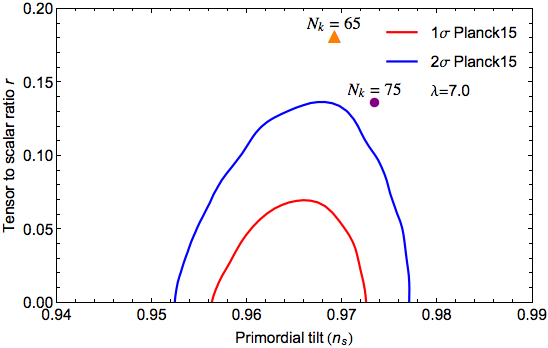}
 \caption{We compare the theoretical predictions in the $(r-n_{s})$ plane for different values of $\lambda$ with Planck$'15$ results for TT, TE, EE, +lowP and assuming $\Lambda$CDM + r \cite{Ade:2015lrj}.}
 \label{fig1}
\end{center}
\end{figure*}

In the following, we will express the power spectrum, the spectral index, and the tensor-to-scalar ratio in terms of $e$-folding number $N_k$.  By substituting the expression for $Q_s$ in Eq. (\ref{B16})  into  Eq. (\ref{B28}) and using the relation (\ref{C23}), we then obtain
\begin{align}
 \mathcal{P_R} &= \frac{1}{3\pi F}\left(\frac{H}{m_{pl}}\right)^2\left(\frac{H}{M}\right)^{2\lambda}\frac{1}{(1+\lambda)^2\epsilon_1^2} \nonumber\\&\simeq \frac{1}{12\pi}\left(\frac{M}{m_{pl}}\right)^2\frac{1}{(1+\lambda)^2\epsilon_1^2} \ . \label{C31}
\end{align}
According to the equation (\ref{N_fold}), the quantities $\mathcal{P_R}$, $n_\mathcal{R}$, and $r$ can be written in terms of $N_k$ as
\begin{eqnarray}
 \mathcal{P_R} = \frac{1}{3\pi}\left(\frac{M}{m_{pl}}\right)^2\frac{N_k^2}{(1+\lambda)^2} \ , \label{C32}
\end{eqnarray}
\begin{eqnarray}
 n_\mathcal{R} - 1 = -\frac{2}{N_k} \ , \label{C33}
\end{eqnarray}
and
\begin{eqnarray}
 r = \frac{12(1+\lambda)^2}{N_k^2} \ . \label{C34}
\end{eqnarray}
Note that the spectral index of scalar perturbations  $n_\mathcal{R}$ does not depend on the rainbow parameter.
%%%%%%%%%%%%%%%%%%%%%%%%%%%%%%%%%%%%%%%%%%%%%%%%%%%%%%%%%%%%%%%%%%%%%%%%%%%%%%%%%%%%%%%%%%%%%%%%%%%%%%%%%%%%%%%%%%%%%%%%%%%%%%%%%%%%%%%%%
\section{Fitting 2015 Planck data} \label{data}
\begin{figure*}
\begin{center}
 \includegraphics[width=0.42\linewidth]{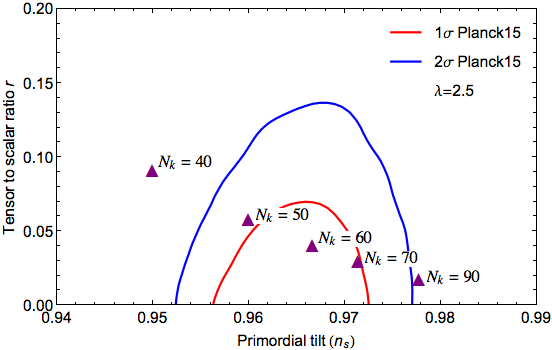}
\includegraphics[width=0.42\linewidth]{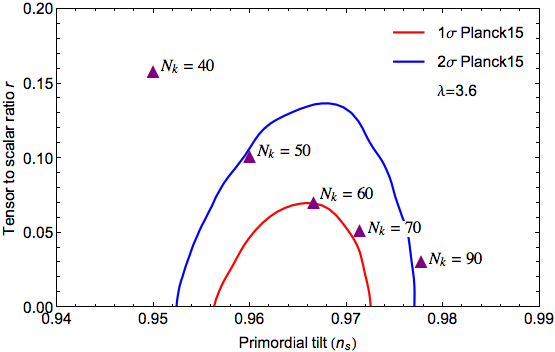}
\includegraphics[width=0.42\linewidth]{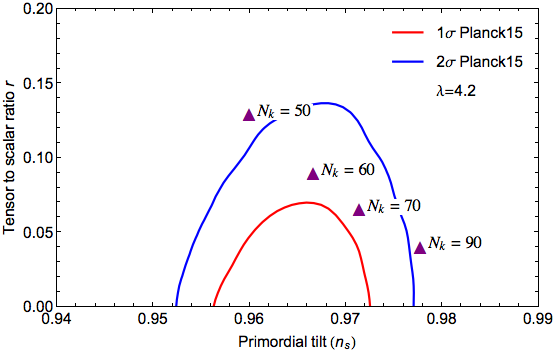}
\includegraphics[width=0.42\linewidth]{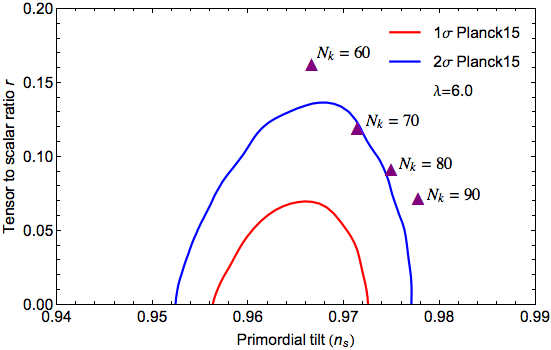}
\includegraphics[width=0.42\linewidth]{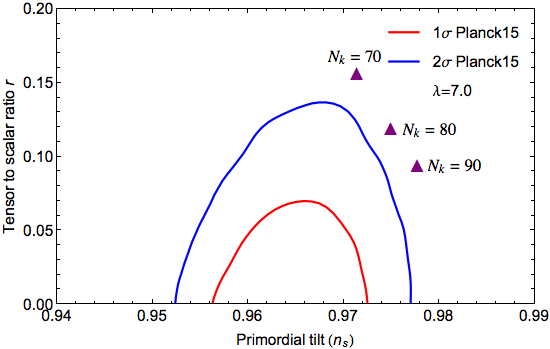}
 \caption{We compare the theoretical predictions in the $(r-n_{s})$ plane for different values of $N_{k}$ for each $\lambda$ with Planck$'15$ results for TT, TE, EE, +lowP and assuming $\Lambda$CDM + r \cite{Ade:2015lrj}.}
 \label{fig2}
\end{center}
\end{figure*}

In this examination, we confront the results predicted by our model with Planck 2015 data. In order to be consistent with Planck data, we find that the value of $\lambda$ cannot be arbitrary. For concreteness, Fig.(\ref{fig1}) and Fig.(\ref{fig2})\,show that to be within $1\sigma$\,C.L.\,of Planck'15 contours value of $\lambda$ cannot be larger than $3.6$ with the sizable number of {\it e}-folds. We find for $\lambda = 3.6$ and $N_{k}=60-65$ {\it e}-folds that the predictions lie well on the boundary of $1\sigma$\,region of the Planck data. Moreover, we discover that the predictions are consistent with the Planck data up to $2\sigma$\,C.L. if the value of $\lambda$ satisfies $\lambda\lesssim 6.0$. Given $\lambda = 6.0$ and $N_{k}=66-72$ {\it e}-folds, the predictions lie on the boundary of $2\sigma$\,region of the Planck data.

\begin{figure*}
\begin{center}
\includegraphics[width=0.42\linewidth]{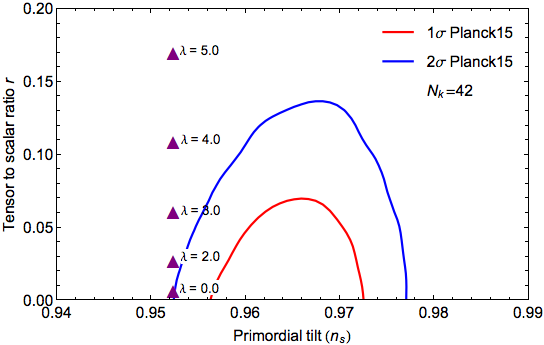}
 \includegraphics[width=0.42\linewidth]{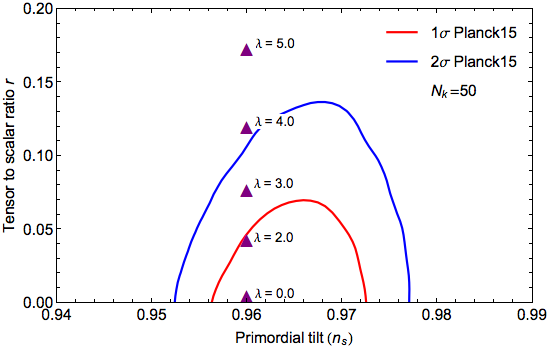}
\includegraphics[width=0.42\linewidth]{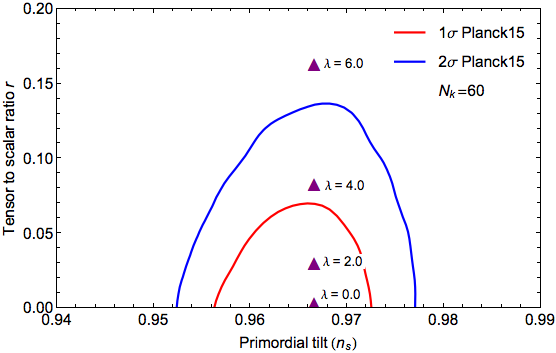}
\includegraphics[width=0.42\linewidth]{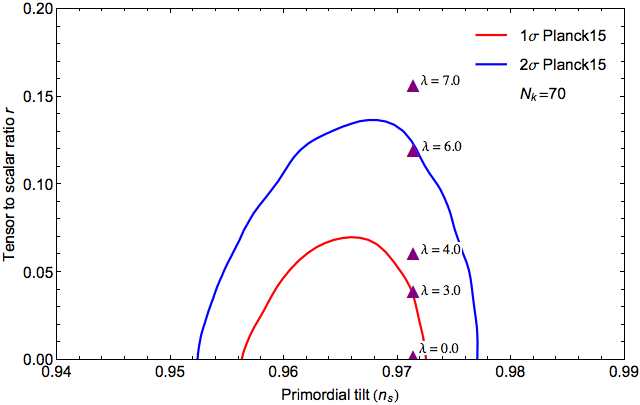}
\includegraphics[width=0.42\linewidth]{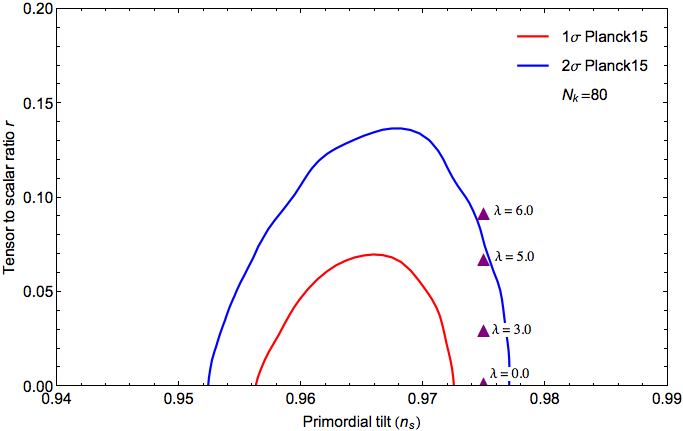}
\includegraphics[width=0.42\linewidth]{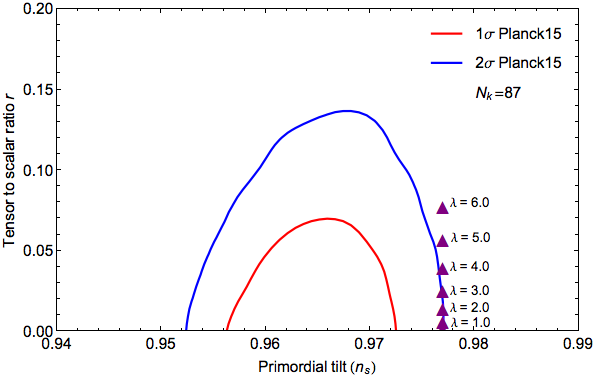}
 \caption{We compare the theoretical predictions in the $(r-n_{s})$ plane for different values of $\lambda$ for each $N_{k}$ with Planck$'15$ results for TT, TE, EE, +lowP and assuming $\Lambda$CDM + r \cite{Ade:2015lrj}.}
 \label{fig3}
\end{center}
\end{figure*}

We also consider the situation in which the values of $\lambda$ are arbitrary but keep $N_{k}$ fixed. In this case, we obtain from Fig.(\ref{fig3})  the window of viable values of $N_{k}$\,{\it e}-folds providing our predictions consistent with the Planck contours. Surprisingly, we discover that in order to be consistent with Planck data up to $2\sigma$\,C.L., the values of $N_{k}$\,{\it e}-folds would satisfy $42\lesssim N_{k}\lesssim 87$ with a proper choice of $\lambda$. However, the values of $\lambda$ cannot be larger than $6.0$.

In some realistic scenario, one can apply $N_{k}=70$ \cite{De_Felice_10}. In this specific case, we find that our predictions nicely satisfy the Planck data at $1\sigma$\,C.L.\, when we opt $\lambda\lesssim 3.0$. Note an extreme (and highly implausible) situation where the number of {\it e}-folds can be even higher, possibly up to 100 \cite{Liddle:2003as}.

Moreover, using parameters of the base $\Lambda$CDM cosmology reported by Planck 2015 \cite{Ade:2015lrj} for $\mathcal{P_R}$ at the scale $k=0.05\,{\rm Mpc^{-1}}$ and taking the value $N_{k}=70$, we find from Eq.(\ref{C32}) that the mass $M$ is constrained to be
\begin{align}
M &\simeq 2\times 10^{-6}(1+\lambda)\,m_{pl}\nonumber\\&\sim \left(0.34 - 1.70\right)\times 10^{14}\,\,{\rm GeV}\,, \label{C35}
\end{align}
with the lower value obtained for the reduced Planck mass of $2.44\times 10^{18}\,\,{\rm GeV}$ and the higher one for the standard one $1.22\times 10^{19}\,\,{\rm GeV}$.

%%%%%%%%%%%CONCLUDE%%%%%%%%%
\section{Conclusion}
\label{con}
In this paper, we have studied the Starobinsky model in the framework of rainbow gravity and obtained the equations of motion where the energy of probe particles varies with the cosmological time. By assuming the rainbow function in the power-law form of the Hubble constant, we have derived the inflationary solution to the equations of motion and calculate the spectral index of curvature perturbation and the tensor-to-scalar ratio.  

By comparing the results predicted by our model with Planck 2015 data, we found that the model is in agreement with the data up to $2\sigma$ if the rainbow parameter satisfies $\lambda\lesssim 6.0$ with the sizable number of $N_{k}$ {\it e}-folds associated with $42\lesssim N_{k}\lesssim 87$. We have also used the scalar power spectrum amplitude reported by Planck 2015 to determine the mass $M$ and found, for $\lambda\lesssim 3.0$ assuming $N_{k}=70$, that $M \sim \left(0.34 - 1.70\right)\times 10^{14}\,\,{\rm GeV}$.
%%%%%%%%%%%%%%%%%%%%%%%%%%%

\acknowledgments
We are grateful to Ahpisit Ungkitchanukit and Khamphee Karwan for discussions and helpful suggestions.  P.C. thanks Peeravit Koad for technical assistance.  The work of A.C. and V.Y. is supported by ``CUniverse'' research promotion project by Chulalongkorn University (grant reference CUAASC) and P.C. is financially supported by the Thailand Research Fund (TRF) under the project of the \lq\lq TRF Grant for New Researcher\rq\rq\, with Grant No. TRG5780143. This work is supported in part by the Special Task Force for Activating Research (STAR) Project, Ratchadaphiseksomphot Endowment Fund, Chulalongkorn University.
%%%%%%%%%%%%%%%%%%%%%%%%%%%%%%%%%%%%%%%%%%%%%%%%%%%%%%%%%%%%%%%%%%%%

\end{document}